%
\input phyzzx.tex
\tolerance=1000
\voffset=-0.3cm
\hoffset=1.0cm
\sequentialequations
\def\rl{\rightline}

\def\t1{{\tilde 1}}

\REF\LEN{L. Susskind, hep-th/9309145.}
\REF\HRS{E. Halyo, A. Rajaraman and L. Susskind, hep-th/9605112.}
\REF\CVE{M. Cvetic and A. Tseytlin, Phys. Rev. {\bf D53} (1996) 5619,
hep-th/9512031;
A. Tseytlin, Mod. Phys. Lett. {\bf A 11} (1996) 689; hep-th/9605091;J. Russo,
hep-th/9606031.}
\REF\FOUR{J. Maldacena and A. Strominger, hep-th/9603060; C. Johnson, R. Khuri
and
R. Myers, hep-th/9603061; G. Horowitz, D. Lowe and J. Maldacena,
hep-th/9603195.}
\REF\KT{I. Klebanov and A. Tseytlin, Nucl. Phys. {\bf B475} (1996) 179,
hep-th/9604166.}
\REF\ROT{J. Breckenridge, R. Myers, A. Peet and C. Vafa, hep-th/9602065;
J.
Breckenridge, D. Lowe, R. Myers, A. Peet, A. Strominger and C. Vafa,
hep-th/9603078.}
\REF\DM{S. Das and S. Mathur, hep-th/9606185; hep-th/9607149.}
\REF\HKRS{E. Halyo, B. Kol, A. Rajaraman and L. Susskind, hep-th/9609075.}
\REF\KG{S. Gubser and I. Klebanov, hep-th/9608108; hep-th/9609076.}
\REF\CGKT{C. Callan, S. Gubser, I. Klebanov and A. Tseytlin, hep-th/9610172.}
\REF\EH{E. Halyo, hep-th/9610068.}
\REF\MS{J. Maldacena and L. Susskind, hep-th/9604042.}
\REF\LW{F. Larsen and F. Wilcek, hep-th/9511064.}

\singlespace
\rl{SU-ITP-51}
\rl{\today}
\pagenumber=0
\normalspace
\medskip
\bigskip
\titlestyle{\bf{ Four Dimensional Black Holes and Strings with Rescaled
Tension}}
\smallskip
\author{ Edi Halyo{\footnote*{e--mail address: halyo@dormouse.stanford.edu}}}
\smallskip
\centerline {Department of Physics} 
\centerline{Stanford University} 
\centerline {Stanford, CA 94305}
\smallskip
\vskip 2 cm
\titlestyle{\bf ABSTRACT}

We give a microscopic description of extreme and near--extreme 
Reissner--Nordstrom black holes in four dimensions in terms of fundamental
strings in a background of magnetic five branes and monopoles. The string
oscillator numbers and tension are rescaled due to the Rindler space
background
with the string mass fixed. The entropy of the black holes is reproduced
correctly
by that of the string with the tension rescaling taken into account.

\singlespace
\vskip 0.5cm
\endpage
\normalspace

\centerline{\bf 1. Introduction}

If string theory is a quantum theory of gravity it has to provide a
microscopic
explanation for black hole entropy given by the Bekenstein--Hawking formula
$$S={A \over{4G_N}} \eqno(1)$$
in terms of string degrees of freedom. One would expect that the black hole
entropy arises from the degenaracy of the string states which describe a black
hole with given mass and charges. Naively, this expectation fails due to the
fact that
the black hole and string entropies scale differently with the mass;
$S_{BH} \sim M^2$ whereas $S_{st} \sim M$ in four dimensions. However, in
ref.[\LEN] it was shown that the scaling is the same if one considers the
Rindler energy of a
string in the black hole background and identifies it with the string entropy.
In [\HRS] this result was expanded and it was shown that $S_{BH}$ and $S_{st}$
completely
agree for Schwarzschild black
holes in any dimension.

During the last year, different microscopic descriptions of four dimensional
Neveu--Schwarz (NS) and Ramond--Ramond (RR)
charged Reissner--Nordstrom (RN)
black holes appeared in the literature [\CVE,\FOUR,\KT]. 
A common feature of these is that the essential degrees of freedom of the
black
hole seem to be those of long (D) strings. For extreme and near--extreme black
holes
these degrees of freedom are non or weakly interacting respectively. Other
objects
such as five (D) branes which are present do not play a dynamical role in the
black hole
but simply constitute backgrounds for the string. This picture also applies to
rotating
black holes and emission of low energy neutral, charged and fixed scalars
[\DM,\HKRS,
\KG,\CGKT].

Recently, another microscopic description of five dimensional NS charged RN
black holes
was given in ref. [\EH]. In this paper, we extend the results of ref. [\EH] to
four dimensional
extreme and near--extreme RN black holes. We show that these black holes
can be described by a string with two charges in a background of magnetic five
branes
and monopoles. Due to the background the string oscillator numbers and tension
are 
rescaled [\LW] whereas the mass remains fixed. The mass of the black hole is
given by the sum
of the background and string masses. The black hole entropy is given only by
that of the
string with the tension rescaling taken into account. This is done by
calculating the Rindler
energy of the string in the background of magnetic five branes and monopoles
and identifying it with the entropy of the string.

This paper is organized as follows. In section 2, we review the classical four
dimensional RN black hole solutions. In section 3, we give the microscopic
description of these black holes in terms of a string with two charges in a
background
of magnetic five branes and monopoles. We apply the method of ref. [\EH] to
extreme and 
near--extreme four dimensional RN black holes. Section 4 contains our
conclusions.

\bigskip
\centerline{\bf 2. Classical Reissner--Nordstrom Black Holes in Four
Dimensions}

In this section, we review the solution for the NS charged RN black hole in
four
dimensions. The classical solution to the low energy equations of motion in
type 
II string theory
compactified on $T^6$ is given by the metric $g_{\mu \nu}$, the NS
antisymmetric tensor
$B_{\mu \nu}$  and the dilaton $g^2=e^{2 \phi}$.
The RR three form, the self--dual five form and the RR scalar are set to zero.
Also, the
asymptotic value of the dilaton $\phi$ is taken to be zero.
The classical four dimensional RN black hole metric is given by
$$ds^2=-\chi^{-1/2} \left(1-{r_0 \over r} \right)dt^2+\chi^{1/2}
\left[\left(1-
{r_0 \over r} \right)^{-1}dr^2+r^2d\Omega_2^2 \right] \eqno(2)$$
where
$$\chi=\left(1+{{r_0^2 sinh^2 \alpha}\over r^2} \right) \left(1+{{r_0^2 sinh^2
\beta} \over r^2} \right)
\left(1+{{r_0^2 sinh^2 \gamma} \over r^2} \right) \left(1+{{r_0^2 sinh^2
\delta}\over r^2} \right) 
\eqno(3)$$
The solution is parametrized by eight parameters, $\alpha, \beta,\gamma,
\delta, r_0$ and
the compactified one and four volumes $2 \pi R, 2 \pi R^{\prime}$ and $(2
\pi)^4 V$.
The total energy of the black hole is
$$E={R R^{\prime}V r_0 \over {2g^2 \alpha^{\prime 4}
}}(cosh 2\alpha+ cosh 2\beta + cosh 2 \gamma +cosh 2 \delta) \eqno(4)$$
The entropy of the black hole is found from the area of the horizon using the
Bekenstein--Hawking formula
$$S={A_H \over {4G_4}}={8 \pi R R^{\prime}V r_0^2 \over g^2
\alpha^{\prime 4}}
cosh \alpha~ cosh \beta~
cosh \gamma~ cosh \delta \eqno(5)
$$
where the ten and four dimensional Newton constants are given by $G_{10}=
8 \pi^6 g^2 \alpha^{\prime 4}$ 
and $G_4={G_{10} / {(2\pi)^6R R^{\prime}V}}$.
The RN black hole carries four NS charges
$$\eqalignno{Q_5&={R^{\prime} r_0\over{2 \alpha^{\prime}}} sinh (2 \alpha)
&(6a)\cr
                  Q_1&={V R^{\prime} r_0\over{2 g^2 \alpha^{\prime 3}}}
sinh (2 \beta) &(6b)\cr             
                 n&={R^2 R^{\prime} V r_0\over{2 g^2 \alpha^{\prime 4}}}
sinh (2 \gamma) &(6c)\cr
                  Q_m&={ r_0 \over {2 }} sinh (2 \delta) &(6d)} $$
These are the charges of the black hole under the NS three form $H_3$, its
dual $H_7$, Kaluza--Klein two form coming from the metric and a magnetic
monopole
charge\foot{I thank Arvind Rajaraman for proposing the magnetic monopole as the
fourth
charge.} .  The magnetic monopole is the only object which can be added to the
five dimensional solution to stabilize the new compactified coordinate with
radius $R^{\prime}$.
Note that due to the
fact that we have a RN black hole the scalars in the solution, i.e. the dilaton
$\phi$
and the  compactification volumes $R,R^{\prime},V$ (which are a priori fields)
are constant in space.
The RN condition is $\alpha=\beta=\gamma=\delta$
so that the four contributions to the mass of the black hole are equal.
The extreme limit is obtained by $r_0 \to 0$ and $\alpha, \beta, \gamma, \delta
\to \infty$
with the charges $Q_5,Q_1,n,Q_m$ fixed. The cases with less than four charges
are
obtained by setting the corresponding angles to zero.

The properties of the black hole can be written in a suggestive way if we
trade
the eight parameters $\alpha, \beta,\gamma, \delta, r_0,R,R^{\prime},V$ for
$N_5, \bar{N_5},N_1,\bar {N_1}, n_L,n_R,N_m, \bar{N_m}$ defined by
$$\eqalignno{N_5&={ r_0 R^{\prime}
\over {4 \alpha^{\prime}}}
e^{2 \alpha} &(7a)\cr
                  \bar {N_5}&={r_0 R^{\prime} \over {4 \alpha^{\prime}}}
e^{-2 \alpha} &(7b)\cr
                  N_1&={V R^{\prime} r_0 \over {4g^2 \alpha^{\prime 3}}}
e^{2 \beta} &(7c)\cr
                   \bar {N_1}&={V R^{\prime} r_0 \over {4g^2 \alpha^{\prime
3}}}
e^{-2 \beta} &(7d)\cr              
                   n_L&={R^2 R^{\prime} V r_0 \over {4g^2 \alpha^{\prime 4}}}
e^{2 \gamma} &(7e)\cr
                   n_R&={R^2 R^{\prime} V r_0 \over {4g^2 \alpha^{\prime 4}}}
e^{-2 \gamma} &(7f)\cr
                    N_m&= { r_0 \over{4  R^{\prime}}}
e^{2\delta}&(7g)\cr
                   \bar{N_m}&={ r_0 \over{4  R^{\prime}}}
e^{-2\delta} &(7f)}
$$
In terms of the above numbers, the charges of the black hole are
$Q_5=N_5-\bar{N_5}$,
$Q_1=N_1-\bar{N_1}$, $n=n_L-{n_R}, Q_m=N_m-\bar{N_m}$. The black hole mass is
$$M_{BH}={RV \over {g^2 \alpha^{\prime 3}}}(N_5+ \bar{N_5})+
{R \over {\alpha^{\prime}}} (N_1+\bar{N_1}) + {1 \over R}(n_L+n_R) + {R
R^{\prime 2} V \over
{g^2 \alpha^{\prime 4}}}
\eqno(8)$$
The entropy can be written as
$$S=2\pi(\sqrt{N_1}+\sqrt{\bar{N_1}}) (\sqrt{N_5}+\sqrt{\bar{N_5}})
(\sqrt{n_L}+ 
\sqrt{n_R}) (\sqrt{N_m}+\sqrt{\bar{N_m}})  \eqno(9)$$
The extreme limit is given by $\bar{N_5}=\bar{N_1}=n_R=\bar{N_m}=0$. For small
deviations beyond
extremality  $\bar{N_5} \sim \bar{N_1} \sim n_R \sim \bar{N_m}<<N_5 
\sim N_1 \sim n_L \sim N_m$.
Note that all three anti--charges
must become nonzero in order to satisfy the RN nature of the
nonextreme black hole. 

Our aim in the next section will be to show that the RN black hole mass and
entropy
are given by that of a fundamental closed string with a rescaled tension in the
background
of magnetic five branes and monopoles. We will identify $Q_1$ and $n$ with the
net 
winding and momentum 
number
of the string whereas $Q_5$ and $Q_m$ will be the net five brane (winding)  and
magnetic 
monopole numbers respectively. 

\bigskip
\centerline{\bf 3. Black Holes and Strings with Rescaled Tension}

In this section, we show that four dimensional RN black holes can be described
by
strings with two charges in a background of  magnetic five branes and
monopoles. The mass of the black hole is given by the sum of the masses of the
background
and the string.  The black hole entropy is given only by that of the string
which is related to the Rindler energy of the string in the background.
The background of  $Q_5$ magnetic five branes and $Q_m$ monopoles is Rindler
space
near the horizon. We calculate the Rindler energy of the string in this
background
and relate it to 
its entropy. We find that this equals the entropy of the black hole; therefore
we
conclude that the string carries 
all of the black hole entropy. We show that this method works for the extreme
and near--extreme 
four dimensional black holes.

{\it 3.1. Extreme Reissner--Nordstrom Black Holes}

In this section we consider extreme RN black holes with four charges. From eqs.
(4) 
and (5) they have 
mass 
(since $\bar{N_5}=\bar{N_1}=n_R=N_m=0$)
$$M_{BH}={RV \over {g^2 \alpha^{\prime 3}}}N_5+
{R \over \alpha^{\prime}}N_1 + {n_L \over R}+ {R R^{\prime 2}V \over {g^2
\alpha^{\prime 4}}}
N_m \eqno(10)$$
and entropy 
$$S=2\pi\sqrt{N_1 N_5 n_L N_m} \eqno(11)$$
For the RN black hole the four contributions to the mass are equal.
Following the arguments of ref. [\EH], we will describe this black hole as a
string with
two charges in a background of $N_5$ magnetic five branes and $N_m$ magnetic
monopoles. 
Two of the charges of the black hole, $N_1$ and $n_L$
are the winding and momentum numbers of the string so that the mass of the
string is
$$M^2_{st}=Q_R^2+{{4N_R} \over \alpha^{\prime}}=Q_L^2+{{4N_L} \over
\alpha^{\prime}} \eqno(12)$$
with 
$$Q_{R,L}=\left((N_1-\bar{N_1}){R \over \alpha^{\prime}} \pm {(n_L-n_R) \over
R} \right) \eqno(13)$$
where the oscillator numbers
are given by
$$N_{L,R}={1 \over 4}[(\sqrt{N_1}+\sqrt{\bar{N_1}})(\sqrt{n_L}+\sqrt{n_R}) \pm
  (\sqrt{N_1}-\sqrt{\bar{N_1}})(\sqrt{n_L}-\sqrt{n_R})]^2 \eqno(14)$$
Note that $N_{L,R}$ above satisfy the level matching condition for any $N_1,
\bar{N_1},n_L,n_R$.
Using the definitions of $N_{L,R}$  we find that for the extreme black
hole
with $n_R=\bar{N_1}=0$,
$N_R=0$ and $N_L=N_1 n_L$.
Therefore the string is in a BPS state. This is not surprising since we are
considering 
an extreme RN black hole which is described by a BPS string state on a BPS
background.
The mass of the background with the string is
$$M={RV \over {g^2 \alpha^{\prime 3}}}N_5+{R R^{\prime 2}V \over g^2
\alpha^{\prime 4}}N_m
+\sqrt{{4N_1 n_L \over \alpha^{\prime}}} \eqno(15)$$
which is equal to that of the black hole given by eq. (10). 
It is easy to show that the lowest excitations
of the string with rescaled tension have energy $\sim 1/(RN_5 N_1  N_m)$
which is precisely that
expected for the four dimensional black hole.
 
The black hole entropy is more difficult to obtain however. Naively, 
the entropy of the black hole is given by that of the string
$$S=2\pi \sqrt{c \over 6} (\sqrt{N_L}+\sqrt{N_R}) \eqno(16)$$
Using the expressions for $N_{L,R}$, we see that we do not get the correct
black hole 
entropy in eq. (11). This is not surprising since
the above formula for entropy holds for a free string whereas we have
a string in a background of five
branes and monopoles.
In the curved background of a black hole
the Rindler energy of an object is given by $dE_R=dM/2 \pi T_R$. Comparing this
with
 $dM=TdS$ for the string we find that
$$S=2\pi \left(T_R \over T \right) E_R \eqno(17)$$ 
The Rindler energy is a dimensionless quantity which should be
identified with the (square root of the)
oscillator number of the string and not with its mass. This is also apparent
from eq. (17) and the expression for the string entropy.

We now compute the Rindler energy of the above string in the background of
$Q_5$ five branes and $Q_m$ monopoles.
The metric
for this background is obtained from eq. (2) by taking $\alpha=\gamma=0$ and
is
given by
$$ds^2=-\chi^{-1/2} \left(1-{r_0 \over r} \right)dt^2+\chi^{1/2}
\left[\left(1-
{r_0 \over r} \right)^{-1}dr^2+r^2d\Omega_2^2 \right] \eqno(18)$$
where
$$\chi=\left(1+{{r_0 sinh^2 \alpha} \over r} \right)  \left(1+{{r_0 sinh^2
\delta} \over r} \right) 
\eqno(19)$$

Near horizon, the metric in eq. (18) gives the Rindler space--time. In this
limit
$$r \to r_0, \quad \chi \to(1+sinh^2\alpha) (1+sinh^2\delta)
=cosh^2\alpha~ cosh^2\delta  \eqno(20)$$
We rescale the metric by
$$r^{\prime}=r \chi^{1/4}, \quad r_0^{\prime}=r_0  \chi^{1/4} \eqno(21)$$
The charges of the background are 
$$Q_5=N_5={r_0 R^{\prime}\over{ \alpha^{\prime}}}sinh(2\alpha) \simeq {2r_0
R^{\prime}\over
{ \alpha^{\prime}}}cosh^2\alpha \eqno(22)$$
and
$$Q_m=N_m={r_0 \over{R^{\prime}}}sinh(2\delta) \simeq {2r_0 \over
{ R^{\prime}}}cosh^2\delta \eqno(23)$$
where the second equalities hold in the near extreme case when
$r_0 \to 0, \alpha, \delta
\to  \infty$ , i.e. $\bar{N_5}=\bar{N_m}=0$.
Then
$$
{\chi}^{1/4} r_0^{\prime}=
r_0 cosh \alpha~ cosh \delta= {1 \over 2} \sqrt{N_5 N_m \alpha^{\prime}}
\eqno(24)
$$
Expanding near the horizon,
$r^{\prime}=r_0^{\prime}+y$ the metric becomes
$$ds^2=-{\chi}^{-1/2}{y\over r_0^{\prime}}dt^2+{r_0^{\prime} \over {y}}dy^2+
r_0^{\prime2}d\Omega_3^2 \eqno(25)$$
The proper distance $\rho$ to the horizon is
$$\rho=\int\sqrt{ r_0^{\prime} \over{y}} dy=2 \sqrt{r_0^{\prime}}\sqrt y
\eqno(26)$$
Then the coefficient of $dt^2$ becomes
$$g_{00}=-{\rho^2 \over{4 {\chi}^{1/2} r_0^{\prime2}}} \eqno(27)$$
One can bring the metric to Rindler form by the rescaling
$$\tau={t \over{2 {\chi}^{1/4} r_0^{\prime}}} \eqno(28)$$
where $\tau$ is the Rindler time conjugate to Rindler energy $E_R$ which is
given by
$$E_R=2 {\chi}^{1/4} r_0^{\prime}M=M\sqrt{N_5 N_m \alpha^{\prime}} \eqno(29)$$
As a result of eq. (17) we identify in the Rindler space--time
$T=2T_R$ and
$$E_R=2\sqrt{c \over 6}\left(\sqrt{N_L^{\prime}}+\sqrt{N_R^{\prime}}\right)
\eqno(30)$$
Now for a BPS state of free string with no charge $M^2=4 N_L/
\alpha^{\prime}$. 
We see that, in the background of
five branes and monopoles, the string oscillator number $N_L$ is rescaled by a
factor of
$\sqrt {N_5 N_m}$
compared to the free string (for $c=6$). 
On the other hand, the mass of the free string (added to that
of the background) gives  the
correct black hole mass. This means that we need to keep the string mass fixed
but
rescale $N_{L,R}$. This  can be done if we assume that  
$\alpha^{\prime}$ (or the string tension) is rescaled
simultaneously with $N_{L,R}$ so that the mass remains the same, i.e.
$\alpha^{\prime} \to
\alpha^{\prime}_{eff}=N_5 N_m \alpha^{\prime}$.  
The entropy of  the string with the rescaled tension is 
obtained 
from eqs. (17) and (30) for $c=6$
$$S=2\pi \sqrt{N_5 N_1 n_L N_m}  \eqno (31) 
$$
which is exactly the entropy of the black hole given by eq. (11).
Note that for a string whose
tension is rescaled as above $T=2T_R$.

We found that due to the presence of the background, the string tension is
rescaled
together with $N_{L,R}$ so that the mass remains the same. Taking 
this rescaling into account
we find that the string entropy is precisely the black hole entropy. Here we
assume that
all gravitational effects which are present in the black hole
are summed up in the rescaling of the string tension. In addition, we assume
that
the masses of the background and the string are additive. This makes sense
since
both the background and the string are BPS states.
Note also that the formula
holds only for $c=6$ whereas for a physical type II string $c=12$. This means
that the
string is effectively confined to the the five brane world--volume and
therefore 
oscillates only in the four transverse
directions.

{\it 3.2 Near--extreme Reissner--Nordstrom Black Holes}

We now consider slightly nonextreme RN black holes with mass
$$
M_{BH}={RV \over {g^2 \alpha^{\prime 3}}}(N_5+ \bar{N_5})+
{R  \over \alpha^{\prime}}(N_1+\bar{N_1}) + {1 \over R}(n_L+n_R)+
{R R^{\prime 2}V \over g^2 \alpha^{\prime 4}} (N_m+ \bar{N_m})
  \eqno (32)
$$
and entropy
$$S=2 \pi (\sqrt{N_1}+\sqrt{\bar{N_1}}) (\sqrt{N_5}+\sqrt{\bar{N_5}})
(\sqrt{n_L}+ 
\sqrt{n_R}) (\sqrt{N_m}+\sqrt{\bar{N_m}})    \eqno(33)$$
where $N_1 \sim N_5 \sim n_L \sim N_m>>\bar{N_1} \sim \bar {N_5} \sim n_R \sim
\bar {N_m}$. 
Note that the deviation from
extremality is for the four charges simultaneously due to the RN nature of the
black hole.
Since the black hole is
nonextreme it is described by a non BPS string state ($N_R \not=0$) in a
nonextreme background (five branes and anti five branes plus monopoles and anti
monopoles). 
The string mass is given by
$$M^2_{st}=Q_R^2+{{4N_R} \over \alpha^{\prime}}=Q_L^2+ {{4 N_L} \over
\alpha^{\prime}} \eqno(34)$$
with 
$$
Q_{R,L}=\left((N_1-\bar{N_1}){R \over \alpha^{\prime}} \pm {(n_L-n_R) \over R}
\right) \eqno(35)
$$
with $N_{L,R}$ given by eq. (14). Using the condition for the extreme RN case
(which holds
approximately for the nonextreme case) 
$${{N_1 R} \over \alpha^{\prime}} \simeq {n_L \over R}  \eqno(36)$$
we find that the string mass is
$$M_{st}=\sqrt{{{4N_1 n_L} \over \alpha^{\prime}}+{{8 N_1 n_R} \over
\alpha^{\prime}}} \eqno(37)$$
For anti--charges much smaller than the charges we find that 
$$M_{st} \simeq {{2 N_1  R} \over \alpha^{\prime}} +{2 n_R \over R}
\eqno(38)$$
Adding this to the mass of the background we obtain
$$M={RV \over {g^2 \alpha^{\prime 3}}}(N_5+ \bar{N_5})+{R R^{\prime 2}V \over
g^2 
\alpha^{\prime 4}}(N_m+\bar{N_m})+{{2 N_1R} \over  \alpha^{\prime}}
  + {2 n_R \over R} \eqno(39)$$
which is the mass of the near--extreme black hole given by eq. (32). 

In order to find the entropy, we have to calculate the Rindler energy of the
above string
$E_R$ in the background of
$N_5$ five branes, $\bar {N_5}$ anti five branes, $N_m$ monopoles and
$\bar{N_m}$
anti monopoles with $N_5, N_m>>\bar {N_5},\bar{N_m}$. Once again
we find 
$$E_R=2M {\chi}^{1/4}r_0^{\prime}  \sim  {Mr_0 } cosh \alpha~ cosh \delta
\eqno (40)$$
in the near extreme limit. But now,
$$Q_5=N_5-\bar{N_5}={r_0 \over \alpha^{\prime}} sinh^2 (2 \alpha) \eqno(41)$$
and
$$Q_m=N_m-\bar{N_m}={r_0 \over \alpha^{\prime}} sinh^2 (2 \delta) \eqno(42)$$
so that the oscillator numbers and the string tension are rescaled by the
factor
$(\sqrt{N_5}+\sqrt{\bar{N_5}})(\sqrt{N_m}+\sqrt{\bar{N_m}})$. The entropy of
the string is 
$$S=2 \pi (\sqrt{N_L^{\prime}}+\sqrt{N_R^{\prime}}) \eqno(43)$$
where $N_{L,R}^{\prime}=
(\sqrt{N_5}+\sqrt{\bar{N_5}})(\sqrt{N_m}+\sqrt{\bar{N_m}})N_{L,R}$. Using the
definition of $N_{L,R}$
in eq. (14) we find the entropy of the string in the background
$$S=2\pi(\sqrt{N_5}+\sqrt{\bar{N_5}}) (\sqrt{N_1}+\sqrt{\bar{N_1}})
(\sqrt{n_L}+ 
\sqrt{n_R}) (\sqrt{N_m}+\sqrt{\bar{N_m}}) \eqno(44)$$
which is the entropy of the near--extreme black hole. The rescaling factor of
the tension
$(\sqrt{N_5}+\sqrt{\bar{N_5}})(\sqrt{N_m}+\sqrt{\bar{N_m}})$ is exactly the
factor needed to convert the string entropy into the
black hole entropy.
We see that even in the near--extreme case the masses of
the background and the string are additive. This does not make sense since
these are not BPS states.

\bigskip
\centerline{\bf 4. Conclusions}

We have shown that extreme and near--extreme RN black holes in
four dimensions can be described by strings with two charges in a background
of magnetic five branes and monopoles. The extreme (nonextreme) black 
holes correspond to BPS (non BPS) states of the string in a BPS (non BPS)
background.
The mass of the black hole is given
by the sum of the string and background masses.
Due to the background,
the string oscillator numbers and tension are rescaled whereas the mass
remains
fixed. The black hole entropy is reproduced only by that of the string with
tension
rescaling taken into account. The string entropy is given by its Rindler
energy
in the background five branes and monopoles.

In this paper as in ref. [\EH] we made a number of assumptions which we are
not
able justify. The main assumption is that all gravitational effects in the
black
hole can be summed up in the rescaling of the string oscillator numbers and
tension. In addition, for both BPS and non BPS states of the string (which 
correspond to 
extreme and near--extreme black holes), the mass remains
fixed in the Rindler background i.e. the masses of the constituents are
additive.
Contrary to the BPS case this does not 
make sense for non BPS states.
Finally, we neglected all possible effects
due to the string mass on the string itself. Since for the RN black holes the
mass of the string equals that of the background, this is difficult to
understand.

\bigskip
\centerline{\bf Acknowledgments}
We would like to thank A. Rajaraman and L. Susskind for useful discussions.

\vfill

\refout
\end

\bigskip
\centerline{\bf Acknowledgments}
We would like to thank M. Dine for very useful discussions.

\end